  \documentclass{aa}

\usepackage{graphicx}
\usepackage{txfonts}
\newcommand{\Arhp}  {\mbox{ArH$^{+}$}}       
\newcommand{\tArhp}  {\mbox{$^{36}$ArH$^{+}$}}       
\newcommand{\HH}  {\mbox{H$_2$}}       
\newcommand{\Ohp}  {\mbox{OH$^{+}$}}       

\newcommand{\ai}{\textit{ab initio}}
\newcommand{\Ai}{\textit{Ab initio}}
\begin{document}

   \title{Photodissociation of interstellar ArH$^+$}

      \author{E. Roueff
          \inst{1}
          \and
            A. B. Alekseyev\inst{2}
           \and    J.  Le Bourlot \inst{1}
          }

   \institute{LERMA and UMR 8112, Observatoire de Paris, Place J. Janssen,
                        92190 Meudon, France\\
              \email{evelyne.roueff@obspm.fr}
         \and
            Fachbereich C, Physikalische und Theoretische Chemie, Bergische Universit\"at Wuppertal,
              Gau\ss strasse 20, D-42097 Wuppertal, Germany\\
               \email{alexeev@uni-wuppertal.de}
                  \and
                   LERMA and UMR 8112, Observatoire de Paris, Place J. Janssen,
                        92190 Meudon, France\\
              \email{jacques.lebourlot@obspm.fr}
              }

     \date{Received  ; accepted  }


  \abstract
   {}
   {Following the recent detection of  {\tArhp} in the Crab nebula spectrum 
   , we have computed the photodissociation 
    rate of {\Arhp} in order to constrain the physical processes at work in this environment.}
   {Photodissociation cross sections of {\Arhp}  are computed in an {\ai} approach including explicit account of spin-orbit coupling.}
   { We report the photodissociation cross section of {\Arhp} as a function of wavelength. Photodissociation probabilities are  derived for different impinging radiation fields.
 The photodissociation probability of for a very small unshielded cloud surrounded on all sides  
 by the unshielded InterStellar 
     Radiation Field (ISRF) model described by  \cite{draine:78} is equal to $9.9 \cdot10^{-12}$ s$^{-1}$  and 1.9 $\cdot 10^{-9}$ s$^{-1}$ in the Crab nebula conditions. The dependence on the visual extinction 
     is obtained by using the Meudon Photon Dominated Region (PDR) code and corresponding analytical fits
     are provided.}
   { These data will help to produce a realistic chemical network to interpret the observations. Photodissociation of  {\Arhp} is found to be moderate and the presence of this molecular ion is mainly dependent on the molecular fraction.}

   \keywords{Molecules --
                Interstellar -medium --
                Molecular Physics
               }

   \maketitle
%

\section{Introduction}

Whereas {\Arhp} has been studied in great detail from laboratory and theoretical techniques, the detection
of the {\tArhp} isotopologue  in the Crab nebula \citep{barlow:13} offers an outstanding opportunity to exploit these studies in a 
previously unforeseen  astrophysical context. The main motivation for this article is to investigate the 
response of this molecular ion to the high ambient radiation field and subsequently check the chemical 
processes at work  in the Crab Nebula environment. 
Section~\ref{sec:theory} describes the recent {\ai} studies of  {\Arhp}.  We use the available information 
to compute photodissociation rates resulting from different shapes of the incident radiation fields as 
discussed in Section~\ref{sec:pdr}. We first describe the main features of Photon Dominated Regions (PDR) 
models  and introduce different radiation fields to compute the photodissociation probabilities.
We summarize our conclusions in Section~\ref{sec:conclusion}.


\section{{\Ai} study of \Arhp}
\label{sec:theory}
 The {\it ab initio} data which are used in the present work to analyze the ArH$^+$ 
photodissociation have been obtained in the detailed theoretical study \citep{alekseyev:07} 
dealing with the computation of molecular potentials and transition moments of the ArH$^+$ 
cation. The multireference spin-orbit CI (Configuration Interaction) method in its $\Lambda-S$ contracted 
version (LSC--SO--CI) has been used for this purpose. It combines a multireference 
single- and double-excitation CI approach (MRD-CI) with relativistic effective core  
potentials (RECP). The LSC--SO--CI method has been described comprehensively in 
\cite{alekseyev:03}, while technical details of the ArH$^+$ computations have been given in 
\cite{alekseyev:07}.

The ground X $^1 \Sigma^+$ state of ArH$^+$ correlates to the Ar($^1S$) + H$^+$ 
asymptote, whereas all low-lying excited electronic states converge to the 
charge-transfer Ar$^+$($^2P^o$) + H($^2S$) limit.  The computed spectroscopic 
constants for the $X\,^1\Sigma^+$ ground state as well as its dipole moment are found 
to be in very good agreement with experiments \citep{johns:84,surkus:00}.  While the ArH$^+$ ground state is well 
bound due to the strong argon-proton interaction, the lowest excited states,  
$b\,^3\Pi$, $B\,^1\Pi$ (...$\sigma^2\pi^3\sigma^*$), $a\,^3\Sigma^+$, and 
$A\,^1\Sigma^+$ (...$\sigma\pi^4\sigma^*$) are all repulsive in the Franck--Condon 
region. They possess very shallow potential minima at large internuclear distances 
due to the weak attraction between the Ar$^+$ cation and hydrogen. The {\it ab initio} 
calculations give for the depths of these wells 230--280 cm$^{-1}$ for most of the
triplet states  and 60--90 cm$^{-1}$ in the case of A$^1\Sigma^+$  and b$^3\Pi_0^+$.

Computing photodissociation of the ArH$^+$ system in realistic conditions requires 
taking into account the spin-orbit (SO) interaction in the Ar$^+$ cation, in which 
the $^2P_{1/2} - {^2P}_{3/2}$ splitting is equal to 1431 cm$^{-1}$ \citep{saloman:10}, 
corresponding to an energy of 2060~K.  
Together with the various potential energy curves converging to the fine structure 
levels of Ar$^+$, \cite{alekseyev:07}  have computed electric dipole moments for 
transitions to the states responsible for the first absorption continuum ($A$ band) 
of ArH$^+$ for the first time in the approach including SO coupling, as well as the 
corresponding extinction coefficients $\varepsilon$ expressed in 
l cm$^{-1}$  mole$^{-1}$. It has been shown that the ArH$^+$ absorption in the $A$ band 
may be divided into two regions. Its high-energy part peaking at 123270 cm$^{-1}$ is dominated by the parallel 
$A\,^1\Sigma^+ \gets X\,^1\Sigma^+$ transition, by far the strongest among transitions 
in this energy range. The low-energy part of the band peaking at 86560 cm$^{-1}$
mainly originates from the perpendicular $B\,^1\Pi \gets X\,^1\Sigma^+$ transition, 
which is $\sim 240$ times weaker than $A \gets X$ at its maximum.  The two partial 
absorption spectra overlap (equal intensities at around $95 \cdot 10^3$ cm$^{-1}$
corresponding to 1053 $\AA$), 
which leads to a shoulder in the red part of the absorption curve. All other 
transitions in this energy region are spin-forbidden, but it can be noted that 
absorption into the $b\,^3\Pi_{0^+}$ state is not negligible in the $(70-80)\cdot 
10^3$ cm$^{-1}$ range. It should be possible to detect it from the product angular 
distribution in the Ar$^+$($^2P_{3/2}$) + H($^2S$) dissociation channel.
The branching ratio $\Gamma$ for the final photodissociation products has also been  
calculated and it has been shown that it smoothly increases from 0 in the red tail 
of the band to 1 at $E \geq 10^5$ cm$^{-1}$.  The latter value corresponds to 
the exclusive formation of the spin-excited Ar$^+(^2P_{1/2})$ ions, that can lead 
to the IR laser generation on the Ar$^+$($^2P_{1/2} - ^2P_{3/2}$) transition.
All spectral calculations have been carried out for $^{40}$ArH$^+$. It is easy to 
show, however, that the zero-point-energy in the ground state of the $^{36}$ArH$^+$
cation increases by $\sim 2$ cm$^{-1}$ with respect to $^{40}$ArH$^+$, which is 
of negligible importance for the present analysis.

In this study, we derive the corresponding photodissociation cross sections from 
the relation $\sigma = 1000 \times log_{10}(e) \times \varepsilon / N_A$.
{$\varepsilon$} is the extinction coefficient computed in \cite{alekseyev:07} expressed in
l mol$^{-1}$ cm$^{-1}$ and N$_A$ is the Avogadro number.
We thus derive $\sigma =  3.82 \times 10^{-21}  \times \varepsilon$, where $\sigma$ is now in $cm^{2}$.  
Figure 1 reports the dependence of the dissociation cross section as a function of 
the wavelength, expressed in \AA. Since all electronically excited potential curves 
are repulsive, the cross section is continuous and does not display any resonance. 
The vertical blue dotted line shows the wavelength cutoff at 912 {\AA}, which takes 
place in astrophysical media, as all photons are absorbed if their energy is above 
the ionization potential of atomic hydrogen.
The threshold of the total photodissociation cross section occurs at 1500 $\AA$. The values display a maximum at 800 $\AA$ and 
become negligible below 670 $\AA$. 
 
   \begin{figure*}
   \centering
  \includegraphics[width=9cm]{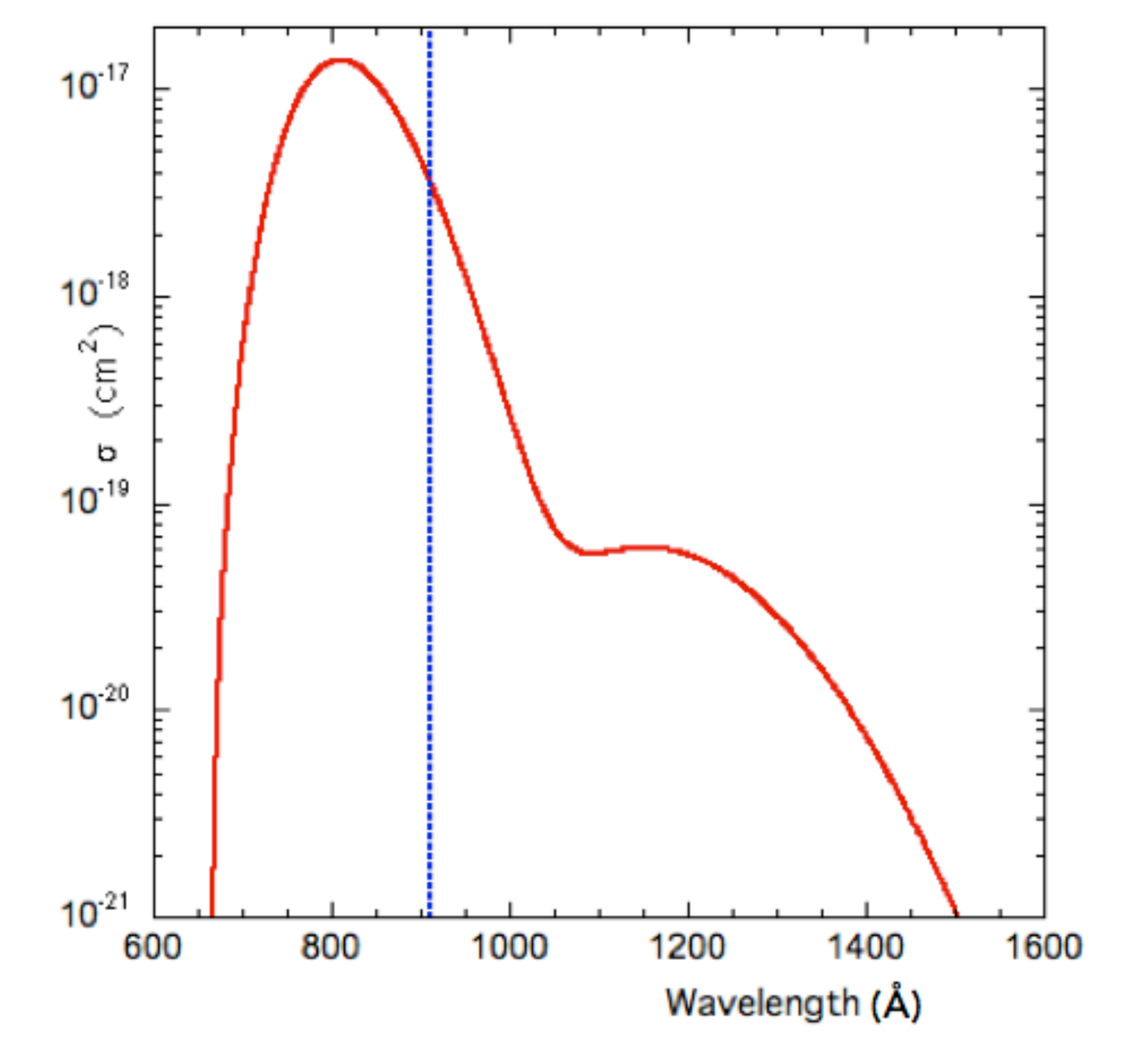}
   \caption{Total photodissociation cross section of {\Arhp} as a function of incident wavelength in $\AA$. The vertical blue dotted line 
   represents the wavelength cutoff of the interstellar radiation field}
              \label{fig:1}%
    \end{figure*}
 We report  the total photodissociation cross section as a function of wavelength  in the online material. 
\onllongtab{
\begin{longtable}{cc}
\caption{Photodissociation cross section of  {\Arhp} as a function of wavelength in $\AA$.}\\
\hline  \hline
Wavelength & Photodissociation cross section   \\  
$\AA$  &  cm$^2$\\
\hline
\endfirsthead
\caption{Continued.}\\
\hline
Wavelength & Photodissociation cross section   \\  
$\AA$  &  cm$^2$\\
\hline
\endhead
\hline
\endfoot
\hline
\endlastfoot
 6.667e+02 &  6.880e-22 \\
  6.681e+02& 1.595e-21   \\
  6.690e+02 & 2.629e-21  \\
6.702e+02 &  4.370e-21  \\
  6.711e+02 &  6.165e-21  \\
  6.720e+02&  8.366e-21  \\
  6.740e+02 &  1.499e-20  \\
  6.790e+02 &   4.382e-20  \\
  6.802e+02  &  5.348e-20  \\
  6.851e+02   &  1.109e-19  \\
6.900e+02     &  2.033e-19  \\
  6.951e+02   &  3.410e-19  \\
6.970e+02  &  4.080e-19  \\
 6.980e+02  &   4.446e-19  \\
  7.002e+02  &   5.363e-19  \\
  7.041e+02  &   7.334e-19  \\
  7.081e+02  &   9.772e-19  \\
  7.101e+02   &  1.117e-18  \\
  7.121e+02   &  1.271e-18  \\
 7.149e+02    &  1.505e-18  \\
 7.170e+02   &  1.695e-18  \\
  7.180e+02   &   1.796e-18  \\
    7.201e+02  &    2.009e-18  \\
  7.230e+02   &    2.328e-18  \\
7.240e+02   &     2.450e-18  \\
7.280e+02   &     2.942e-18  \\
  7.330e+02  &     3.645e-18  \\
  7.352e+02  &     3.973e-18  \\
    7.390e+02   &  4.580e-18  \\
  7.401e+02  &   4.760e-18  \\
  7.420e+02  &  5.081e-18  \\
  7.450e+02  &   5.596e-18  \\
  7.470e+02  &   5.936e-18  \\
 7.490e+02   &    6.284e-18  \\
  7.501e+02    &  6.486e-18  \\
  7.520e+02  &   6.849e-18  \\
  7.555e+02  &   7.480e-18  \\
  7.560e+02  &   7.583e-18  \\
  7.580e+02  &   7.949e-18  \\
  7.630e+02  &   8.824e-18  \\
  7.700e+02  &   1.005e-17  \\
  7.760e+02  &  1.101e-17  \\
  7.790e+02   &  1.144e-17  \\
  7.820e+02    &  1.184e-17  \\
  7.870e+02    &  1.243e-17  \\
  7.910e+02   &   1.286e-17  \\
  7.970e+02  &   1.332e-17  \\
  8.002e+02   &   1.348e-17  \\
 8.102e+02   &    1.373e-17  \\
  8.145e+02   &   1.370e-17  \\
  8.192e+02    &  1.355e-17  \\
  8.260e+02    &   1.318e-17  \\
8.304e+02     &    1.287e-17  \\
8.360e+02     &     1.238e-17  \\
8.402e+02     &     1.193e-17  \\
  8.452e+02   &     1.136e-17  \\
  8.502e+02   &  1.076e-17  \\
  8.560e+02    &  1.002e-17  \\
 8.590e+02   &     9.634e-18  \\
  8.702e+02   &    8.148e-18  \\
 8.751e+02   &      7.518e-18  \\
  8.860e+02   &     6.154e-18  \\
  8.903e+02   &    5.650e-18  \\
  8.951e+02   &    5.126e-18  \\
  8.991e+02   &    4.710e-18  \\
  9.003e+02    &    4.588e-18  \\
  9.052e+02    &  4.110e-18  \\
  9.060e+02   &    4.034e-18  \\
  9.081e+02   &   3.843e-18  \\
  9.102e+02   &   3.660e-18  \\
  9.106e+02    &    3.624e-18  \\
  9.110e+02	&	3.589e-18   \\
 9.202e+02	&	2.871e-18  \\
 9.301e+02	&	2.229e-18     \\
 9.401e+02	&	1.696e-18     \\
 9.504e+02	&	1.271e-18    \\
 9.600e+02	&	9.565e-19\\
 9.702e+02	&	7.010e-19\\
 9.802e+02	&	5.149e-19\\
 9.904e+02	&	3.723e-19\\
 9.953e+02	&	3.187e-19\\
 1.000e+03	&	2.729e-19\\
 1.005e+03	&	2.338e-19\\
 1.010e+03	&	2.033e-19\\
 1.015e+03	&	1.743e-19\\
 1.020e+03	&	1.523e-19\\
 1.025e+03	&	1.319e-19\\
 1.030e+03	&	1.167e-19      \\
 1.035e+03	&	1.026e-19    \\
 1.040e+03	&	9.214e-20\\
 1.045e+03	&	8.351e-20\\
 1.050e+03	&	7.651e-20\\
 1.055e+03	&	7.105e-20\\
 1.060e+03	&	6.681e-20\\
 1.063e+03	&	6.456e-20\\
 1.067e+03	&	6.238e-20\\
 1.070e+03	&	6.108e-20\\
 1.073e+03	&	6.001e-20\\
 1.077e+03	&	5.887e-20\\
 1.080e+03	&	5.826e-20\\
 1.083e+03	&	5.776e-20\\
 1.087e+03	&	5.738e-20\\
 1.090e+03	&	5.722e-20\\
 1.093e+03	&	5.715e-20\\
 1.097e+03	&	5.719e-20\\
 1.100e+03	&	5.730e-20\\
 1.110e+03	&	5.791e-20\\
 1.120e+03	&	5.883e-20\\
 1.130e+03	&	5.974e-20\\
 1.140e+03	&	6.036e-20\\
 1.150e+03	&	6.078e-20\\
 1.160e+03	&	6.081e-20\\
 1.170e+03	&	6.032e-20\\
 1.180e+03	&	5.948e-20\\
 1.190e+03	&	5.814e-20\\
 1.200e+03	&	5.646e-20\\
 1.220e+03	&	5.210e-20\\
 1.240e+03	&	4.679e-20\\
 1.260e+03	&	4.076e-20\\
 1.280e+03	&	3.453e-20\\
 1.299e+03	&	2.880e-20\\
1.300e+03	&	2.855e-20\\
 1.320e+03	&	2.308e-20\\
 1.340e+03	&	1.806e-20\\
 1.360e+03	&	1.384e-20\\
 1.380e+03	&	1.041e-20\\
 1.400e+03	&	7.560e-21\\
 1.420e+03	&	5.398e-21\\
 1.441e+03	&	3.721e-21\\
 1.461e+03	&	2.519e-21\\
 1.490e+03	&	1.361e-21\\
 1.510e+03	&	8.606e-22\\
 1.521e+03	&	6.746e-22\\
 1.530e+03	&	5.386e-22\\
 1.550e+03	&	3.234e-22\\
 1.571e+03	&	1.862e-22\\
 1.591e+03	&	1.059e-22\\
 1.599e+03	&	7.594e-23\\
  \hline
\end{longtable}
}
 \section{Photodissociation probabilities}
  \label{sec:pdr}
We now consider  {\Arhp} photodissociation for different impinging radiation fields. As emphasized in \cite{vd:06},   
photodissociation occurs in the surface layers of molecular environments exposed to UV radiation fields. We use the 
Meudon PDR code facility\footnote{available at http://pdr.obspm.fr} \citep{lepetit:06} to compute the photodissociation 
probability of  {\Arhp} for different incident radiation fields as a function of the visual magnitude.  

 The main feature of  the Meudon PDR code is a detailed computation of the radiative transfer of the VUV radiation 
field in a cloud model containing both gas and dust. Photodissociation of molecular hydrogen is computed from 
detailed molecular properties of Lyman and Werner transitions as computed by \cite{abgrall:93,abgrall:00} involving 
a discrete absorption followed by  emission towards the continuum of the X molecular state. Absorption by   gas and 
dust is treated in a consistent way and we refer the reader to the reference paper describing the Meudon PDR code 
\citep{goicoechea:07,lepetit:06}.  
  \subsection{Photodissociation in an unshielded environment.}
  \label{sec:free}
 
\begin{figure*}
   \centering
 \includegraphics[width=10cm]{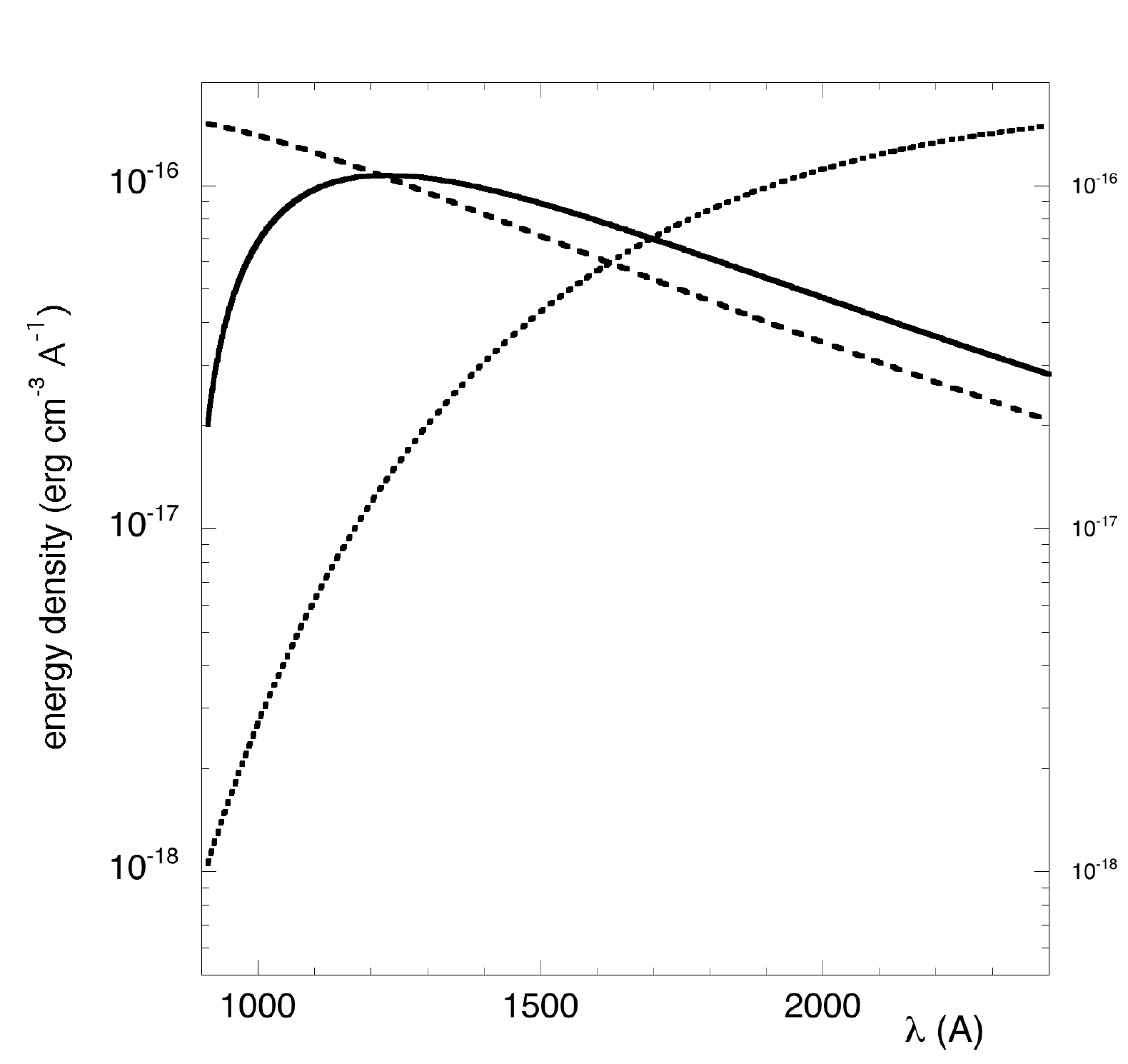} 
   \caption{Energy density of the  radiation fields as a function of wavelength. 
            Solid line: Draine, dotted line:  black body at T=10,000K, 
            dashed line: black body at T=37,000K.}
              \label{fig:2}%
    \end{figure*}

 The photodissociation (photoionization) probability of any molecular   (atomic) species 
 for a very small parcel of gas surrounded on all sides by the ISRF is expressed as
\begin{equation}  \label{eq1}
 k_{pd}^{cont} =  \int_{912}^{\lambda max} \sigma(\lambda)~ N(\lambda) ~d\lambda\ , 
 \end{equation}
where N($\lambda$) is the number of photons per square centimeter per second and per $\AA$
and $ \sigma(\lambda)$ the corresponding cross section.
 We express this quantity as a function of  the radiation energy density via
\begin{equation} \label{eq2}
  N(\lambda)= \frac{\lambda}{4 \pi h} u(\lambda),
\end{equation}
which is valid for an isotropic incident radiation field.  We show in Figure \ref{fig:2} the incident energy 
densities of three radiation fields which have been used for the computations.  We have constrained those 
radiation fields to contain the same incident integrated energy density for a wavelength range between 912 
and 2400 $\AA$, corresponding to the Habing cutoff of the UV radiation field \citep{habing:68}.  This choice 
is slightly different from that made by \cite{vd:06}, who have used a cutoff of 2050 $\AA$  \footnote{We have verified that we obtain identical results for carbon photoionization 
in that  case.}. 
The model using the incident Draine radiation field is 
defined as the standard case.
\\
The radiation field impinging the molecular knots where {\Arhp} has been detected \citep{barlow:13} may 
be estimated as well. The pulsar wind nebula (PWN) luminosity  is obtained from a dust + photoionization modeling
of the Crab nebula by Owen and Barlow (2014)\footnote{Owen P., Barlow M.J., in preparation} using the spectrum displayed in \cite{hester:08}. The integrated PWN luminosity is 1.3 $\times$ 10$^{38}$ 
 erg/s and is assumed to be emitted from the central two-thirds of the ellipsoidal nebula, whose overall dimensions are adopted to be 3.8pc x 2.9 pc.  
Table \ref{tab:res} displays the  photodissociation rate (P$_d$)  of ArH$^+$ as well as the photoionization 
probability (P$_i$) of atomic carbon computed from the direct integration in equation \ref{eq1}
for the three chosen radiation fields and  for the crab nebula environment.
We also display the 
ratios of the radiation field energy densities to that corresponding to the Draine radiation field in the 911 - 2400 $\AA$ window. 
 The results  concerning photoionization of carbon given by \cite{vd:06}  are also reported. 
 The differences obtained reflect the different choice of normalization of the radiation fields. 
 The value displayed for the Crab nebula is about two hundred times larger than in standard interstellar conditions
 and the ratio of the corresponding energy densities is about 80.
 \begin{table*}[h]
 \caption{Photodissociation rates computed for an unshielded environment. VD06 refers to \cite{vd:06}.}             
 \label{tab:res}      
 \centering                          
 \begin{tabular}{ | l | l | c | c c | }        
 \hline\hline                 
 Radiation field & ratio of energy  &Pd($\Arhp$) (s$^{-1}$)  & \multicolumn{2}{c}{Pi(C)  (s$^{-1}$)}   \\    
   &         densities           &       &               present result & VD06\\
 \hline                        
 Draine &    1  &9.9 $\times$ 10$^{-12}$  & 3.1 $\times$ 10$^{-10}$ & 3.1 $\times$ 10$^{-10}$      \\      
 Black body of 10000K &   1   &  7.0 $\times$ 10$^{-13}$& 1.7  $\times$ 10$^{-11}$&   2.5 $\times$ 10$^{-11}$   \\
 Black body of 37000K &  1     & 3.7  $\times$ 10$^{-11}$&   7.6 $\times$ 10$^{-10}$ &    \\
 Crab nebula   & 83.9   & 1.9  $\times$ 10$^{-9}$       &    4.1   $\times$ 10$^{-8}$      &    \\
 \hline                                   
 \end{tabular}
 \end{table*}
%
%
\subsection{Role of the dusty environment}
\begin{figure*}[h]
   \centering
 \includegraphics[width=12cm]{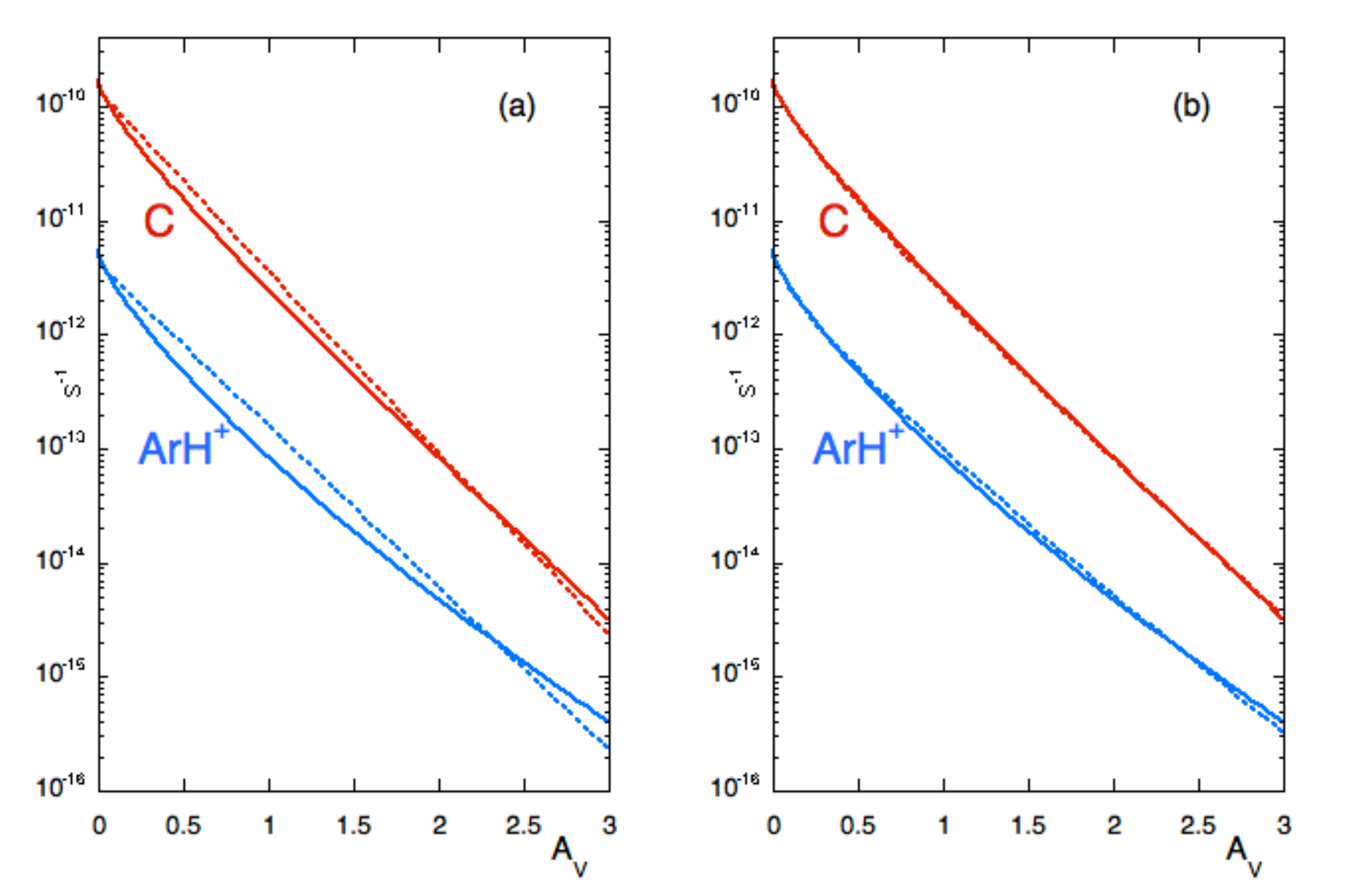} 
   \caption{Computed photodissociation and photoionization rates of  $\Arhp$  (blue) and C (red) at the surface of a semi-infinite slab
   for a Draine isotropic ISRF. Full lines correspond to PDR model results and dotted lines display the analytical fits : (a) exponential;  (b) E$_2$ function.}
              \label{fig:av}%
    \end{figure*}

We consider now a cloud   irradiated  
on one side and we follow the decrease of the radiation field as it penetrates inside the molecular cloud. 
This reduction is a result of photon absorption both by gas and dust.  Dust absorption and scattering properties are 
 introduced, using the prescription by \cite{weingartner:01} for the Milky Way for spherical dust particles with radii 
 between 0.3 $\mu$ and 3 nm following a  -3.5 power   law of  the dust size distribution \citep{mathis:77}.
Dust absorption is a continuum process as well as gas phase absorption by \Arhp, as shown in Figure \ref{fig:1}. 
We also recall that such a continuum  mechanism occurs in the photoionization of carbon as the corresponding cross section is constant from  I$_C$ = 11.26 eV (corresponding to 1101.5 $\AA$), equal to 1.6 $\cdot$ 10$^{-17}$ cm$^2$ up to 13.6 eV. 
An exponential decrease is often used to represent the A$_V$ dependence  of the photoprocesses and
 the corresponding parameters are found in the different kinetic databases (KIDA (/http:://kida.obs.u-bordeaux1.fr) 
\citep{wakelam:12}, UDFA (http://www.udfa.net)  \citep{mcelroy:13}). 
These expressions are aimed to describe the effect of dust continuum absorption on the photoionization and photodissociation probabilities.
We consider a cloud model with a proton density n$_H$ = 10$^3$ cm$^{-3}$ and 
submitted to a cosmic ionization rate of $5\cdot10^{-17}$ s$^{-1}$, corresponding to standard translucent cloud conditions. The radiation,  impinging isotropically on one side of a semi-infinite slab, is described by the Draine analytic  ISRF  model \citep{draine:78}. 
We display as full lines  in Figure \ref{fig:av} the computed 
 {\Arhp} photodissociation and Carbon photoionization probabilities as a function of A$_V$.
The corresponding exponential 
fits are displayed as dotted lines in the left hand part (a) of Figure \ref{fig:av} for a restricted visual extinction range between 0 and 3, as recommended in \cite{vd:06}.   %
We also display in the right hand part (b) of Figure \ref{fig:av}, as dotted lines, the corresponding fitting functions using the  
exponential integral  E$_2$ function.
Whereas the exponential dependence is a fair approximation of the photodissociation probabilities for the considered range of visual extinctions, the fit using the E$_2$ function is superior as already emphasized and used in \cite{neufeld:09}. Corresponding formulae are displayed in Table \ref{tab:fit}.
   
\begin{figure*}[h]
 \centering
 \includegraphics[width= 14cm]{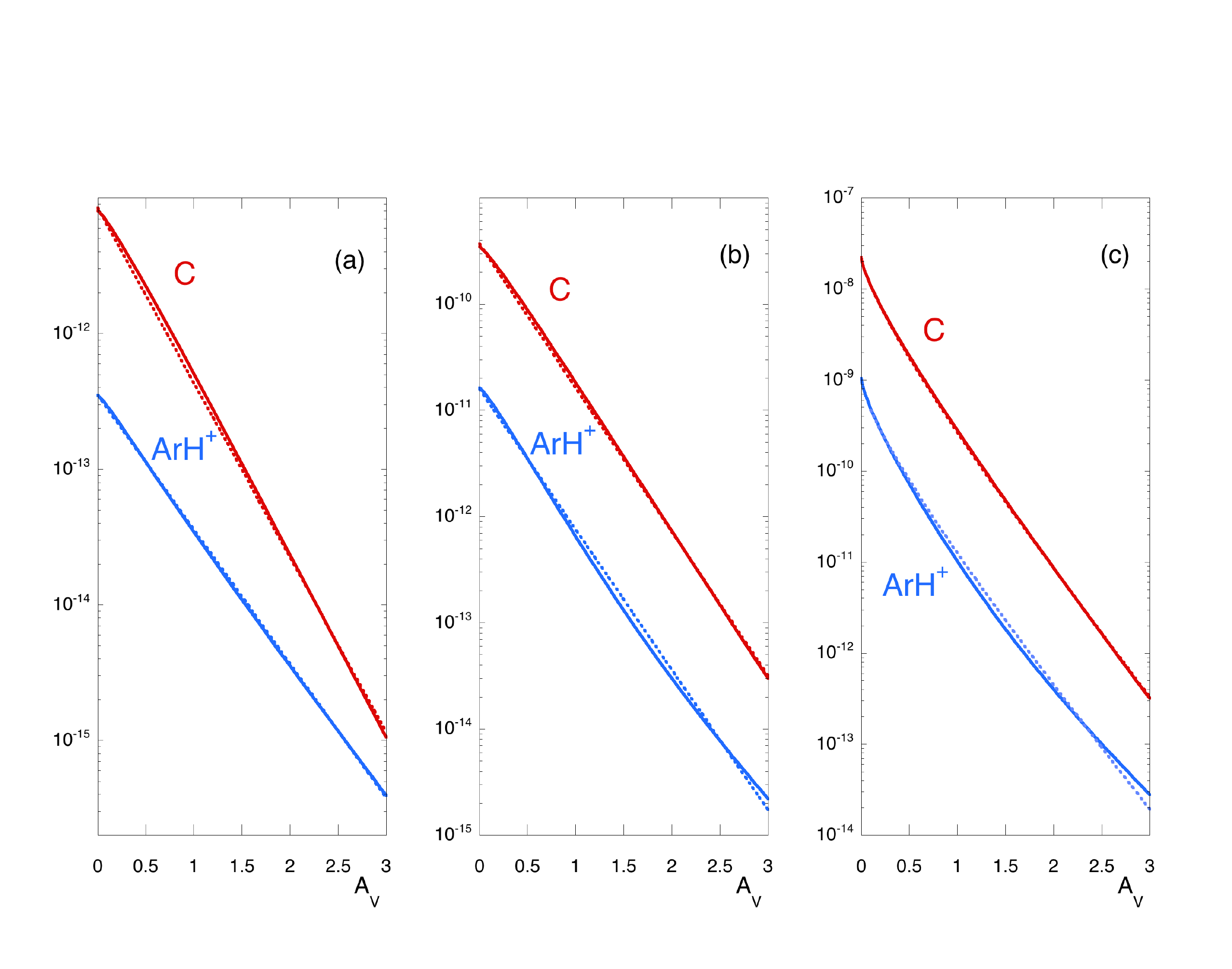} 
   \caption{Computed photodissociation and photoionization rates of  $\Arhp$ (blue) and C (red) at the surface of a semi-infinite slab
    as a function of visual extinction A$_V$.
   Full lines correspond to PDR model results and dotted lines display the fit equations results,  
   as given in Table \ref{tab:fit} . (a) (b)  and (c) represent the BB 10,000~K,  BB 37,000~K and
 the Crab nebula cases. 
  }
              \label{fig:av2}%
    \end{figure*}

 \begin{table*}[h]
\caption{Fitting functions of the photodissociation rate of {\Arhp} and photoionization rate of Carbon as functions of the visual magnitude
for the different radiation fields used in Table \ref{tab:res}. IS and PP stand respectively  for isotropic  and perpendicular irradiation.}             
\label{tab:fit}      
\centering                          
\begin{tabular}{|l|l|l|}
\hline\hline                 
Radiation field & Pd($\Arhp$) (s$^{-1}$)  & Pi(C)  (s$^{-1}$)  \\    
 \hline
 IS  Draine   &  $   4.2 \times 10^{-12}  \times  \exp(-3.27~A_V)$ &   $ 1.40 \times 10^{-10}  \times  \exp(-3.67 ~A_V)$  \\
 IS Draine & $   4.84 \times 10^{-12} \times E_2 (2.46~A_V)$  &   $1.78 \times  10^{-10} \times E_2 (2.84~A_V) $ \\
 PP BB 10,000K &  $  3.54 \times 10^{-13}  \times \exp(-2.28 ~A_V) $&   $ 8.91 \times 10^{-12}  \times  \exp(-2.99 ~A_V)$  \\
 PP BB 37,000K  &  $  1.56 \times 10^{-11}  \times \exp(-3.04~A_V)$ &   $  3.83 \times 10^{-10}  \times  \exp(-3.14~ A_V) $ \\
 IS Crab  &   $9.66  \times 10^{-10} \times E_2 (2.83~A_V)$ &  $2.29  \times 10^{-8} \times E_2 (2.93~A_V)$ \\
 \hline
 \end{tabular}
 \end{table*}
 We also compute the photodissociation rate resulting from radiating stars, considered as the blackbodies already introduced in Section \ref{sec:free} and displayed in Figure \ref{fig:2}, where the radiation is considered as beamed perpendicularly to the surface of the cloud, which corresponds to regions
 located close to a particular star. 
 In those cases, exponential fits are found to be adequate representations of the visual extinction A$_V$ dependence of the photodissociation rates, as shown in Figure \ref{fig:av2}. Finally, we also compute the photodissociation probability corresponding to the Crab nebula conditions where the radiation is isotropic. We assume a proton density 
 n$_H$ = 2 $\times$10$^4$ cm$^{-3}$ representative of the molecular knots detected by \cite{loh:12} in this environment. 
 The corresponding results are displayed in Figure \ref{fig:av2} and fitting expressions are given in Table \ref{tab:fit}.
 The constant factors expressed in s$^{-1}$ are about one half the value given for the unshielded case (Table \ref{tab:res})
 as the radiation is impinging on one single side of the cloud.  

%
\subsection{Role of the continuous gas phase absorption}
   \begin{figure*}[h]
   \centering
 \includegraphics[width=6cm]{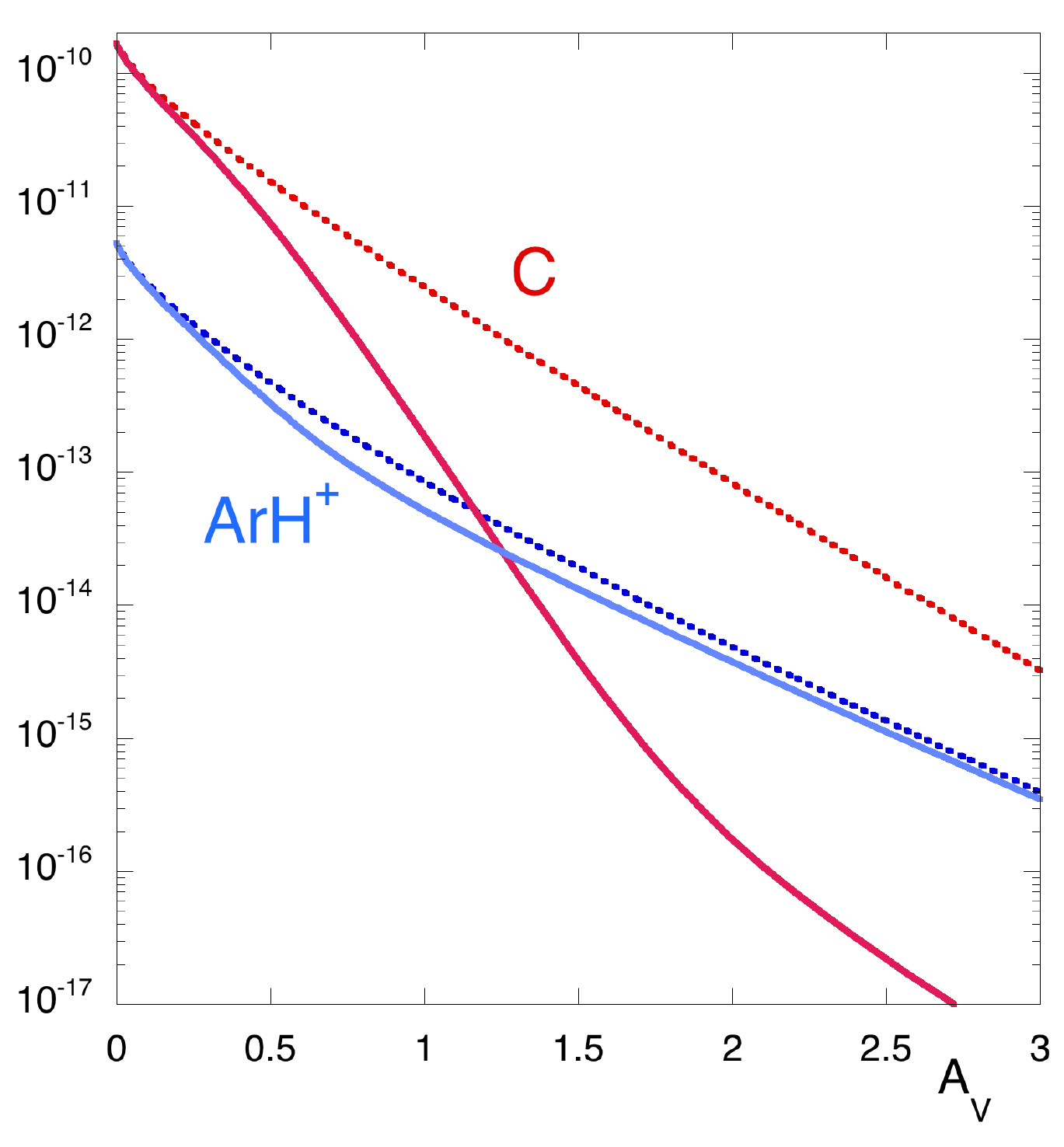} 
   \caption{Computed photodissociation and photoionization rates of  $\Arhp$  (blue curves) and C (red curves) at the surface
   of a semi-infinite slab for the Draine radiation field;
  solid line: PDR model including continuous dust + gas phase absorption; dotted line: PDR model with continuous dust absorption
  only.}
              \label{fig:absgas}%
    \end{figure*}
However, these analytic representations of photodissociation/photoionization probabilities neglect
the role of other gas phase continuum absorptions. Reemphasizing early  PDR models by \cite{tielens:85},  \cite{rollins:12} 
have pointed out recently that neglecting mutual shielding by dust and carbon  leads to overestimate  the photoionization 
probability of Carbon by a factor exp(-$\sigma_C$N$_C$), where $\sigma_C$ is the photoionization cross section of Carbon and N$_C$ stands for the column density of atomic Carbon. 
We investigate in Figure \ref{fig:absgas} the role of additional gas phase absorption for the standard model case, where we introduce the photodissociation and photoionization 
continuous cross sections of different gas phase compounds, including Carbon.  
 The solid line curves correspond to models including continuous absorption of the radiation field by gas whereas dotted lines
 curves correspond to computations neglecting that effect. 
  Whereas gas phase absorption considerably modifies the photoionization probability A$_V$ dependence for Carbon, the effect on {\Arhp} 
  photodissociation is much less significative as the photodissociation threshold is at 1500 $\AA$, at a significantly larger wavelength than for Carbon photoionization (1105 $\AA$).
 \subsection{Relevance to the Crab environment}
\begin{figure*}[h]
   \centering
 \includegraphics[width=7cm]{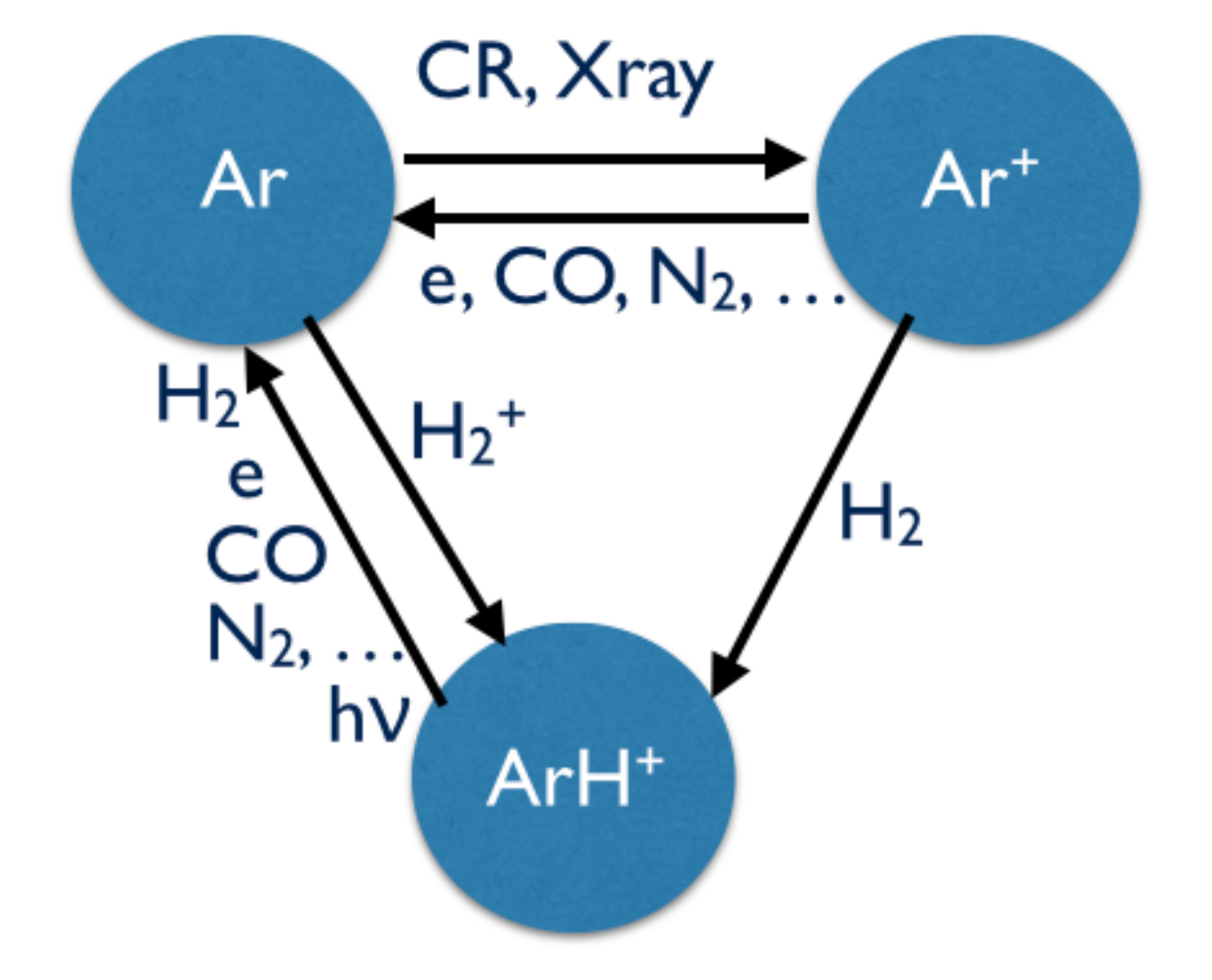} 
   \caption{Chemical network of the {\Arhp} molecular ion.}
              \label{fig:schema}%
    \end{figure*}

 The main chemical processes involved in the formation/destruction of {\Arhp}, except photodissociation,  have been discussed in \cite{barlow:13}.
The availability of the photodissociation cross sections allows to constrain further the chemical processes at work.
The chemical network is relatively simple as displayed in Figure \ref{fig:schema}.
The initial step of the chemistry is ionization of Argon through cosmic rays and X-rays  followed by a reaction with molecular hydrogen.
An alternative path is provided by the reaction between H$_2^+$ with neutral Argon. The abundance  of {\Arhp} is thus directly proportional 
to the  total  ionization rate of  Ar. 
The  ionization state of Ar is discussed by \cite{jenkins:13} in
the context of the Warm Neutral Medium (WNM) and  
non conventional sources of ionization are proposed to reconcile a [ArI / OI]  ratio that is consistent  both with the observations and the generally accepted value of the density n(HI) = 0.5 cm$^{-3}$. 
It is interesting to note that the molecular {\Ohp} ion, which is recognized   to be a test of cosmic ionization rate \citep{indriolo:13}, has been detected in the same environment \citep{barlow:13}. 
 A quantitative study is beyond the scope of the present paper and a realistic study would require the introduction of the X ray spectrum. 
The photodissociation rate of {\Arhp} at the surface of the molecular knots is computed to be  about 2 $\cdot$ 10$^{-9}$ s$^{-1}$   which may compete  with dissociative recombination ( with corresponding rate constant k$_e$ $\le$ 5 $\times$ 10$^{-8}$ cm$^3$ s$^{-1}$ \citep{mitchell:05}).  However reactions with molecular hydrogen (with a rate $\sim$ to 2 $\times$ 10$^{-9}$ cm$^3$ s$^{-1}$ \citep{anicich:03}) are presumably  the main destruction channel. 
 
\section{Summary}
\label{sec:conclusion}
We report photodissociation cross sections of {\Arhp} as a function of wavelength which can be used to compute the response of that molecular ion 
 to ultraviolet interstellar radiation fields.  The photodissociation rate of  {\Arhp} for a very small unshielded  cloud surrounded on all sides by   the standard 
Draine UV ISRF is $9.9\cdot10^{-12}$ s$^{-1}$.  This value is very sensitive to the spectral distribution 
in the VUV 911 - 2400 $\AA$ wavelength window, 
as shown in Table \ref{tab:res}.  
The corresponding rate relevant to  the Crab nebula environment 
is   1.9 $\cdot$ 10$^{-9}$ s$^{-1}$.  
The dependence on the visual extinction A$_V$  due to the attenuation of radiation  by dust 
 is derived from the Meudon PDR code both for incident isotropic and plane parallel radiations. We  report the 
corresponding analytic functions for a semi infinite plane parallel cloud for different typical cases and for the
Crab nebula environment. The constant numerical factor is close to half the value 
obtained for the unshielded environment. The dependence on A$_V$ is better represented with an E$_2$ function when isotropic radiation is impinging on the cloud. Nevertheless, we stress out that a direct integration of the product of
photodissociation cross sections by the actual radiation field allows  to derive photo destruction rates unambiguously.
The actual value of the photodissociation rate is moderate and destruction of  {\Arhp} is mainly due to molecular \HH.
Then, {\Arhp}, as well as {\Ohp} also detected in the Crab nebula \citep{barlow:13}, are strongly dependent on the molecular fraction
of the gas. 
 \begin{acknowledgements}
   We thank M.J. Barlow for various informations on the Crab nebula  environment. 
   After the refereeing process of the paper, we have been aware of an additional detection of {\Arhp} in the diffuse ISM
   \citep{schilke:14}.
   We thank particularly D. Neufeld and J.H. Black for helpful exchanges and related informations. 
   ER and JLB thank  support from the CNRS National Program PCMI (Physico-Chimie du Milieu Interstellaire).
   ABA acknowledges support from the Deutsche Forschungsgemeinschaft.
   \end{acknowledgements}

\bibliography{roueff-arhp}
\bibliographystyle{aa}

\end{document}